\documentclass{jpsj2}

\title{Origin of the Canonical Ensemble: Thermalization with Decoherence}

\author{\textsc{Shengjun Yuan }$^{1,2}$\thanks{E-mail: s.yuan@science.ru.nl}, \textsc{Mikhail I. Katsnelson}$^{1}$, and \textsc{Hans De Raedt}$^{2}$}

\inst{$^{1}$Institute of Molecules and Materials, Radboud University of Nijmegen,
NL-6525ED Nijmegen, The Netherlands \\
$^{2}$Department of Applied Physics, Zernike Institute for Advanced Materials,
University of Groningen, Nijenborgh 4, NL-9747AG Groningen, The Netherlands}

\abst{We solve the time-dependent Schr{\"o}dinger equation for the combination 
of a spin system interacting with a spin bath environment.  
In particular, we focus on the time development of the reduced density matrix of the spin system.  
Under normal circumstances we show that the environment drives the reduced density matrix 
to a fully decoherent state, and furthermore the diagonal elements of the 
reduced density matrix approach those expected for the system in the canonical ensemble.  
We show one exception to the normal case is if the spin system cannot exchange energy with the spin bath. 
Our demonstration does not rely on time-averaging of observables
nor does it assume that the coupling between system and bath is weak.
Our findings show that the canonical ensemble is a state
that may result from pure quantum dynamics, 
suggesting that quantum mechanics may be regarded
as the foundation of quantum statistical mechanics.}

\kword{Quantum Statistical Mechanics, Canonical Ensemble, Time-dependent Schr{\"o}dinger Equation, Thermalization, Decoherence}

\begin{document}
\maketitle


\section{Introduction}

Statistical mechanics is one of cornerstones of
modern physics but its foundations and basic postulates are still under
debate~\cite{balescu,Popescu2006,Rigol2008,Goldstein2006,Reimann2007,Reimann2008,Gemmer2006,Gemmer2006b,
Cazalilla2006,Rigol2006,Rigol2007,Eckstein2008,Cramer2008,Cramer2008b,Flesch2008,Bocchieri1959,Shankar1985,Tasaki1998,sait96,Esposito2003,Merkli2007}.
There is a common believe that a generic 
\textquotedblleft system\textquotedblright\ 
that interacts with a generic environment evolves
into a state described by the canonical ensemble. 
Experience shows that this is true but a detailed understanding of this process, 
which is crucial for a rigorous justification of statistical physics and thermodynamics, 
is still lacking. 
In particular, in this context the meaning of 
\textquotedblleft generic\textquotedblright\ is not clear. 
The key question is to what extent the evolution to the equilibrium state depends on the details of the dynamics of
the whole system. 

Earlier demonstrations that the system can be in the canonical ensemble state
are based on showing that time-averages of the expectation dynamical variables of the system
approach their values for the subsystem that is the thermal equilibrium state~\cite{Bocchieri1959,Shankar1985,Tasaki1998,sait96}
or do not consider the dynamics of the system but
assume that the state of the whole system has a special property called 
``canonical typicality''~\cite{Popescu2006,Rigol2008,Goldstein2006,Reimann2007,Reimann2008,Gemmer2006,Gemmer2006b}
in which case it is as yet unclear under which conditions the
whole system will evolve to the region in Hilbert space where its subsystems
are in the thermal equilibrium state.
A very different setting to study nonequilibrium quantum dynamics 
is to start from an eigenstate of some initial Hamiltonian and
push the system out of this state by a sudden change of the model parameters~\cite{Cazalilla2006,Rigol2006,Rigol2007,Eckstein2008,Cramer2008,Cramer2008b,Flesch2008}.
To the best of our knowledge, it has not yet been shown that this approach 
leads to the establishment of the canonical equilibrium distribution.
Finally, we want to draw attention to the fact that a demonstration
of relaxation to the canonical distribution requires a system
with at least three different eigenenergies 
because a diagonal density matrix of a two-level system can always 
be represented as a canonical distribution~\cite{Esposito2003,Merkli2007}.

The main result of this paper is that we show, without 
any time-averaging procedure or any approximation,
that systems embedded in a closed quantum system 
generally evolve to their canonical distribution states.
This result complies with the fact that if we make a real measurement of a
thermodynamic property, we observe its equilibrium value without having to perform time averaging.
Furthermore, we show that the relaxation to the canonical distribution is not limited
to the regime of weak coupling between system and environment,
an assumption that is often used~\cite{balescu,Popescu2006,Rigol2008,Goldstein2006,Reimann2007,Reimann2008,Gemmer2006,Gemmer2006b}.

\section{General theory}

In general, the state of a closed quantum system is described by a
density matrix~\cite{Neumann55,BALL03}. 
The canonical ensemble is characterized by a density matrix
that is diagonal with respect to the eigenstates of
the system Hamiltonian, the diagonal elements
taking the form $\exp(-\beta E_{i})$ where 
$\beta=1/k_BT$ is proportional to the inverse temperature
($k_B$ is Boltzmann's constant) 
and the $E_{i}$'s denote the eigenenergies.
The time evolution of a closed quantum system is governed
by the time-dependent Schr{\"o}dinger equation (TDSE)~\cite{Neumann55,BALL03}.
If the initial density matrix of an isolated quantum system is non-diagonal,
then, according to the TDSE, its density matrix remains nondiagonal and never
approaches the thermal equilibrium state with the canonical distribution.
Therefore, in order to thermalize the system $S$, it is necessary
to have the system $S$ interact with an environment ($E$), also called heat bath.
Thus, the Hamiltonian of the whole system ($S+E$) takes the form
$H=H_{S}+H_{E}+H_{SE}$, where $H_S$ and $H_E$
are the system and environment Hamiltonian, respectively and  
$H_{SE}$ describes the interaction between the system and environment.

The state of system $S$ is described by the reduced density matrix 
\begin{equation}
\widetilde\rho(t)\equiv\mathbf{Tr}_{E}\rho \left( t\right)
,
\label{eq1}
\end{equation}
where $\rho \left( t\right) $ is the density matrix of the whole system at time $t$
and $\mathbf{Tr}_{E}$ denotes the trace over the degrees of freedom of the environment. 
The system $S$ is in its thermal equilibrium state if the reduced density matrix
takes the form 
\begin{equation}
\widehat\rho\equiv\left.{e^{-\beta H_{S}}}\right/{\mathbf{Tr}_{S}e^{-\beta H_{S}}}
,
\label{eq2}
\end{equation}
where $\mathbf{Tr}_{S}$ denotes the trace over the degrees of freedom of the system $S$.
Therefore, in order to demonstrate that the system $S$, evolving
in time according to the TDSE, relaxes to its thermal equilibrium state
one has to show that $\widetilde\rho\left( t\right)\approx\widehat\rho$
for $t>t_0$ where $t_0$ is some finite time.

The difference between the state $\widetilde\rho\left( t\right)$ 
and the canonical distribution $\widehat\rho$
is most conveniently characterized by the two quantities
$\delta(t)$ and $\sigma (t)$ defined by
\begin{equation}
\delta(t)=\sqrt{\sum_{i=1}^N\left( \widetilde\rho_{ii}(t) -
\left.{e^{-b(t) E_{i}}}\right/{\sum_{i=1}^{N} e^{-b \left( t\right) E_{i}}}\right) ^{2}}
,
\end{equation}%
with 
\begin{equation}
b(t)=\frac{\sum_{i<j,E_{i}\neq E_{j}}
[\ln \widetilde\rho_{ii}(t) -\ln \widetilde\rho _{jj}(t)]/({E_{j}-E_{i}})}{\sum_{i<j,E_{i}\neq E_{j}}1},
\end{equation}%
and%
\begin{equation}
\sigma (t) =\sqrt{\sum_{i=1}^{N-1}\sum_{j=i+1}^{N}\left\vert\widetilde\rho_{ij}(t) \right\vert ^{2}}.
\end{equation}%
Here $N$ denotes the dimension of the Hilbert space of system $S$
and $\widetilde\rho_{ij}(t)$ is the matrix element $(i,j)$
of the reduced density matrix $\widetilde\rho$ in the representation
that diagonalizes $H_S$.
As the system relaxes to its canonical distribution
both $\delta(t)$ and $\sigma (t)$ vanish, $b(t)$ converging
to $\beta$.
As $\sigma(t)$ is a global measure for the size of the 
off-diagonal terms of the reduced density matrix,
$\sigma(t)$ also characterizes the degree of coherence
in the system: If $\sigma(t)=0$ the system is in a state of full decoherence.

\section{Model and simulation method}

To study the evolution to the canonical ensemble state in detail,
we consider a general quantum spin-1/2 model
defined by the Hamiltonians 
\begin{eqnarray}
H_{S} &=&-\sum_{i=1}^{n_{S}-1}\sum_{j=i+1}^{n_{S}}\sum_{\alpha
=x.y,z}J_{i,j}^{\alpha }S_{i}^{\alpha }S_{j}^{\alpha }, \label{HAMS} \\ 
H_{E} &=&-\sum_{i=1}^{n-1}\sum_{j=i+1}^{n}\sum_{\alpha =x,y,z}\Omega
_{i,j}^{\alpha }I_{i}^{\alpha }I_{j}^{\alpha },  \label{HAME}\\ 
H_{SE} &=&-\sum_{i=1}^{n_{S}}\sum_{j=1}^{n}\sum_{\alpha =x,y,z}\Delta
_{i,j}^{\alpha }S_{i}^{\alpha }I_{j}^{\alpha }.  \label{HAMSE}
\end{eqnarray}%
Here the $S^\alpha$'s and $I^\alpha$'s denote the spin-1/2 operators
of the system and environment respectively (we use units such that
$\hbar$ and $k_B$ are one).
Analytic expressions for $\rho(t)$ can only be obtained for very special
choices of the exchange integrals $J_{i,j}^{\alpha }$, $\Omega _{i,j}^{\alpha }$ and $%
\Delta _{i,j}^{\alpha }$ but it is straightforward to 
solve the TDSE numerically for any choice of the model parameters.
Here, we numerically solve the TDSE for $H=H_{S}+H_{E}+H_{SE}$ using the
Chebyshev polynomial algorithm \cite{TALE84,LEFO91,Iitaka97,DOBR03}. 
These \emph{ab initio} simulations yield results that are very accurate
(at least 10 digits), independent of the time step used~\cite{RAED06}.

The state, that is the density matrix $\rho(t)$ of the whole system 
at time $t$ is completely determined by the choice of the initial state of the whole system
and the numerical solution of the TDSE.
In our work, the initial state of the whole system (S+E) is a pure state. 
This state evolves in time according to 
\begin{eqnarray}
|\Psi(t)\rangle&=&e^{-iHt}|\Psi(0)\rangle=\sum_{i=1}^{2^{n_s}} \sum_{p=1}^{2^n} c(i,p,t)|i,p\rangle
,
\label{eq4}
\end{eqnarray}%
where the states $\{ |i,p\rangle \}$ denote a complete set of orthonormal states.
In terms of the expansion coefficients $c(i,p,t)$, the reduced density matrix reads
\begin{eqnarray}
\widetilde\rho(t)_{i,j} &=&\mathbf{Tr}_{E} \sum_{p=1}^{2^n}\sum_{q=1}^{2^n} c^\ast(i,q,t)c(j,p,t)|j,p\rangle\langle i,q|
\nonumber \\
&=&\sum_{p=1}^{2^n} c^\ast(i,p,t)c(j,p,t)
,
\label{eq5}
\end{eqnarray}
which is easy to compute from the solution of the TDSE.
Another quantity of interest that can be extracted from the solution of the TDSE
is the local density of states (LDOS)
\begin{eqnarray}
\mbox{LDOS}(E)&\equiv& \frac{1}{2\pi}\int_{-\infty}^{+\infty} dt\; e^{-iEt} \langle \Psi(0)|e^{-iHt}|\Psi(0)\rangle
\nonumber \\
&=& \sum_{k=1}^{D} |\langle \Psi(0)|\varphi_k\rangle|^2 \delta(E-E_k)
\nonumber \\
&=& 
\sum_{i=1}^{2^{n_s}} \sum_{p=1}^{2^n} c^\ast(i,p,0)c(i,p,t)
,
\label{eq6}
\end{eqnarray}%
where $D=2^{n+n_S}$, $\{|\varphi_k\rangle\}$, and $\{E_k\}$ denote the dimension of the Hilbert space, 
the eigenstates and eigenvalues of the whole system, respectively. 
The LDOS is ``local'' with respect to the initial state: It provides information about
the overlap of the initial state and the eigenstates of $H$.

The notation to specify the initial state is as follows:
$\left\vert GROUND\right\rangle _{S}$ 
is the ground state or a random superposition of all
degenerated ground states of the system;
$\left\vert RANDOM\right\rangle _{S}$ denotes
a random superposition of all possible basis states;
$\left\vert UU\right\rangle _{S}$ is a state in which
all spins of the system are up
meaning that in this state, the expectations value of each spin is one;
$\left\vert UD\right\rangle _{S}$ is a state in which
two nearest-neighbor spins of the system are antiparallel
implying that in this state, the correlation
of their $z$-components is minus one;
and 
$\left\vert RR\right\rangle _{S}$ denotes the product state
of random superpositions of the states of the individual
spins of the system.
The same notation is used for the spins in the environment,
the subscript $S$ being replaced by $E$.

As we report results for many different types of spin systems
it is useful to introduce a simple terminology to classify them according
to symmetry and connectivity.
The terms ``XY'', ``Heisenberg'', ``Heisenberg-type'' and ``Ising''
system refer to the cases $J_{i,j}^{x}=J_{i,j}^{y}=J$ and $J_{i,j}^{z}=0$,
$J_{i,j}^{x}=J_{i,j}^{y}=J_{i,j}^{z}=J$,
$J_{i,j}$ uniform random in the range
$[-\left\vert J\right\vert,\left\vert J\right\vert ]$,
and $J_{i,j}^{x}=J_{i,j}^{y}=0$ and $J_{i,j}^{z}=J$, respectively.
The same terminology of symmetry is used for the Hamiltonian $H_E$
of the environment and for the interaction Hamiltonian $H_{SE}$.
In our model, all the spins of the system interact with each spin of the environment.
To characterize the connectivity of spins within the system (environment), 
we use the term ``ring'' for spins forming 
a one-dimensional chain with nearest-neighbor interactions and
periodic boundary conditions,
``triangular-lattice'' if the spins are located on
a two-dimensional triangular lattice with nearest-neighbor interactions,
and ``maximum-connectivity-system'' when all the spins within the system (environment) interact with each other.

\begin{figure}[t]
\begin{center}
\includegraphics[width=15cm]{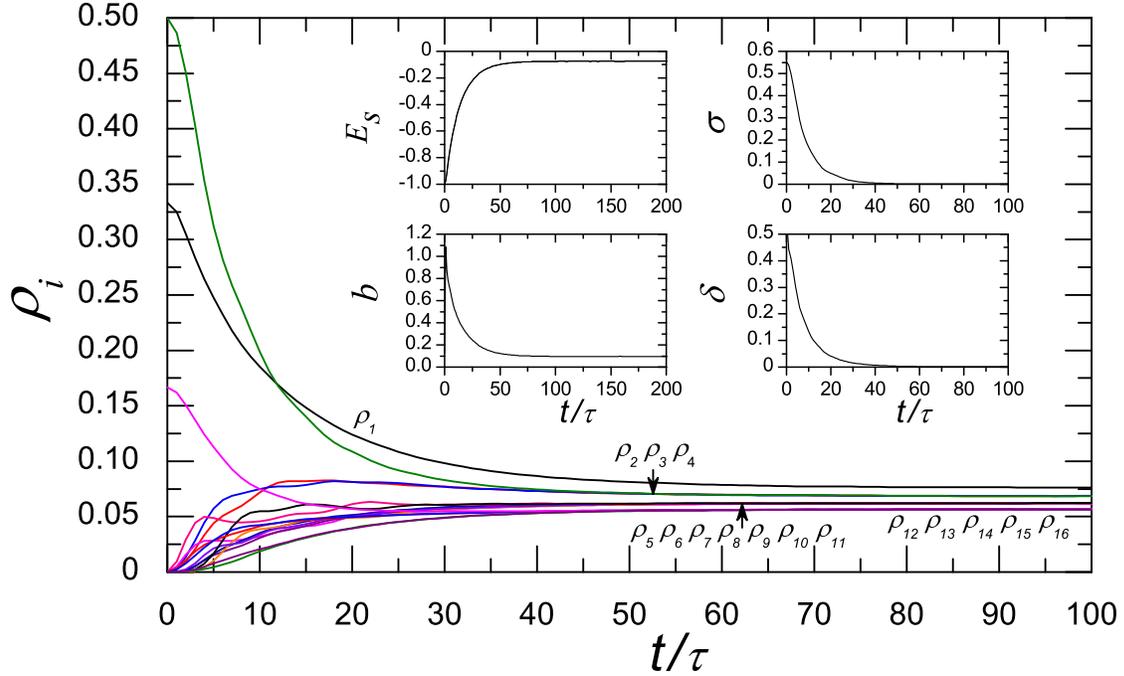} 
\end{center}
\caption {(Color online)
Simulation results for the diagonal elements $\rho_i\equiv\widehat\rho_{ii}(t)$
of the density matrix of $S$, the energy $E_S\equiv E_S(t)$,
the effective inverse temperature $b\equiv b(t)$ and its variance $\delta \equiv \delta(t)$, and 
$\sigma \equiv \sigma(t)$ which
is measure for the decoherence in $S$, as obtained by solving the TDSE for the whole system
with Heisenberg-ring $H_{S}$ ($J=-1$, $n_S=4$), Heisenberg-type $H_{SE}$ ($\Delta =0.3$),
spin glass $H_{E}$ ($\Omega =1$, $n=18$), and $\tau=\pi/10$.
The initial state of the whole system is a product state
of $\left\vert UD\right\rangle _{S}$ and $\left\vert RANDOM\right\rangle_{E}$.
}
\label{fig1}
\end{figure}

\section{Results}

\begin{figure*}[t]
\begin{center}
\includegraphics[width=15cm]{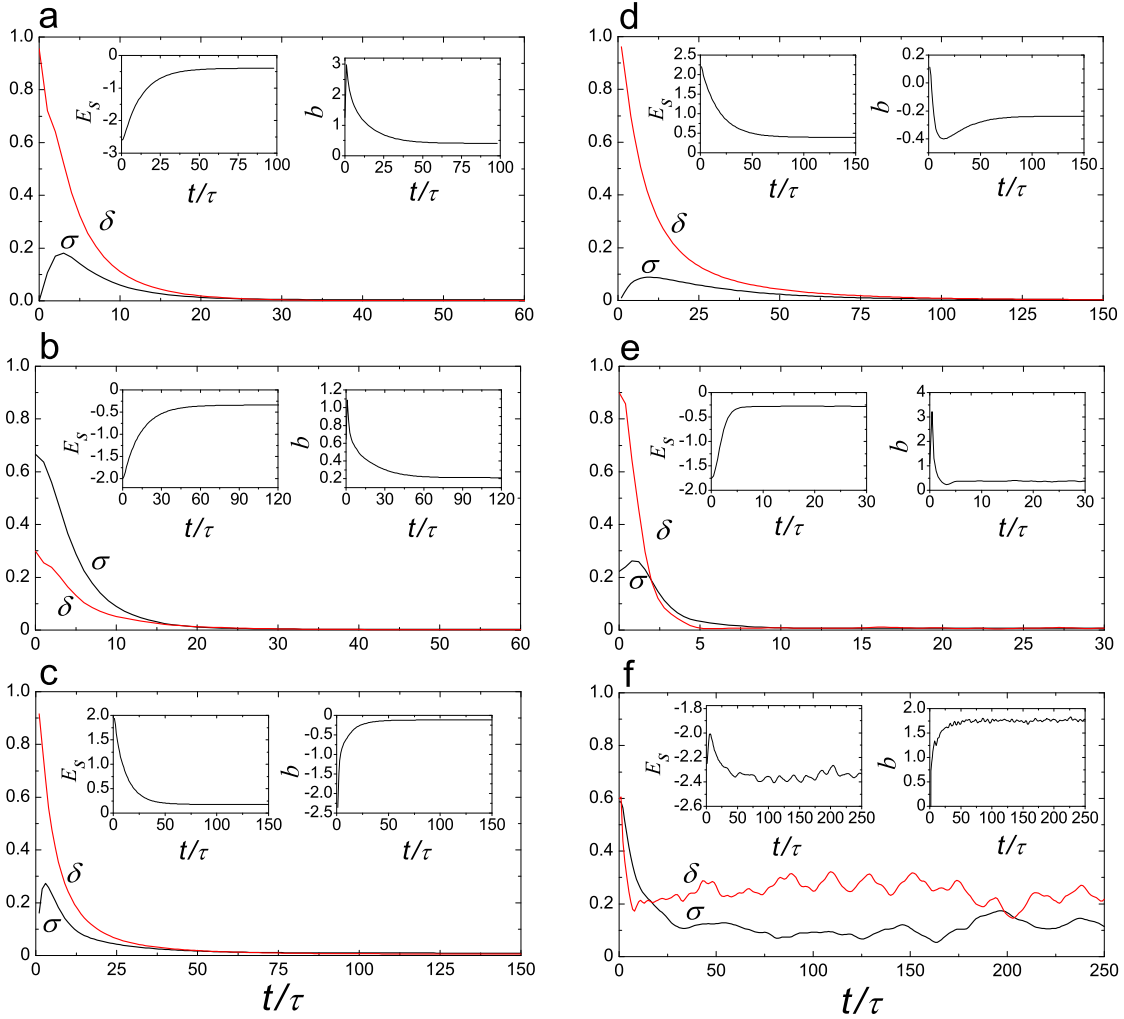}
\end{center}
\caption{ (Color online)
Simulation results for the energy $E_S\equiv E_S(t)$,
the effective inverse temperature $b\equiv b(t)$, its variance $\delta \equiv \delta(t)$, and 
the deviation from a diagonal matrix $\sigma \equiv \sigma(t)$
as obtained by the solution of the TDSE
for a variety of different systems $S$
coupled to a spin glass $H_{E}$ via a Heisenberg-type $H_{SE}$.
The systems used are \textbf{a}: XY-ring, \textbf{b} and \textbf{f}: Heisenberg-ring, \textbf{c}: Ising-ring,
\textbf{d}: Heisenberg-triangular-lattice, and \textbf{e}: spin glass (Heisenberg-type maximum-connectivity-system). 
The initial states of the whole system are
\textbf{a}: $\vert GROUND\rangle_{S}\otimes \vert RANDOM\rangle _{E}$, 
\textbf{b}: $\vert UD\rangle _{S}\otimes \vert RANDOM\rangle _{E}$, 
\textbf{c}: $\vert UU\rangle _{S}\otimes \vert RR\rangle _{E}$, 
\textbf{d}: $\vert UU\rangle _{S}\otimes \vert RANDOM\rangle _{E}$,
\textbf{e}: $\vert GROUND\rangle _{S}\otimes \vert UD\rangle_{E}$, 
and
\textbf{f}: $\vert UD\rangle _{S}\otimes \vert GROUND\rangle _{E}$. 
The numbers of spins in the system are $n_{S}=8$ for cases \textbf{a}-\textbf{c} and $n_{S}=6$ for cases \textbf{d}-\textbf{f}. 
The numbers of spins in the environment is $n=16$ for all cases.
The model parameters are $J=-1$, $\Delta =0.3$ and $\Omega =1$, 
except for case \textbf{e} in which $\Delta =1$.
}
\label{fig2}
\end{figure*}

In earlier work, it was found that a frustrated spin glass (Heisenberg-type maximum-connectivity-system)
environment is very effective for creating full decoherence 
($\sigma \rightarrow0 $) in a two-spin system~\cite{Yuan2006,Yuan2007,Yuan2008}.
As $\sigma \rightarrow0 $ is a necessary condition for the state of 
the system to converge to its canonical distribution,
we have chosen spin glass
environments, which have no obvious symmetries, for further exploration.

First, we consider a system ($H_S$: Heisenberg-ring) interacting 
($H_{SE}$: Heisenberg-type)
with an environment ($H_E$: spin glass). 
The system has four distinct eigenvalues
($E_{1}=-2$, $E_{2-4}=-1$, $E_{5-11}=0$, and $E_{12-16}=1$)
and sixteen different eigenstates. The environment has $2^{18}$ eigenstates. 
During the time-integration of the TDSE, 
the reduced density matrix of the system is calculated every $\tau =\pi /10$.
Following the general procedure described earlier,
the values of the diagonal elements $\widehat\rho_{ii}$
yield an estimate for the effective inverse temperature $b(t)$,
the error $\delta(t)$ for this estimate
and the measure $\sigma(t)$ for the deviation from a non-diagonal matrix.
We also monitor the energy $E_S(t)=\mathbf{Tr}_S\widehat\rho(t)H_S$,
of the system.

From the simulation results, shown in Fig.~\ref{fig1}, 
it is clear that for $t>50\tau$, each diagonal element $\widehat\rho_{ii}$ 
of the reduced density matrix converges to one out of four stationary values,
corresponding to the four non-degenerate energy levels of the system. 
This convergence is a two-step process.
First the system looses all coherence, as indicated
by the vanishing of $\sigma \left( t\right) $ for $t>50\tau$.
The time dependence of $\sigma \left( t\right)$ fits very well to an exponential law
\begin{equation}
\sigma \left( t\right) =\sigma_\infty+Ae^{-t/T_{2}}
,
\end{equation}
with $\sigma_\infty=0.00128$, $A=0.602$ and $T_{2}=8.01\tau$. 
Likewise, the vanishing of $\delta(t)$ on the same time-scale
indicates that the density matrix of the system converges 
to the canonical distribution.
The effective temperature $b(t)$ and the 
energy of the system $E_S\left( t\right)$ also
fit very well to the exponential laws
\begin{equation}
b\left( t\right) =\beta+Be^{-t/T_{1}}
, 
\end{equation}
and
\begin{equation}
E\left( t\right) =E_\infty+Ce^{-t/T_{1}}
,
\end{equation}
with 
$\beta=0.0962$, $B=-0.900$, and $T_{1}=13.3\tau$ and
$E_\infty=-0.0745$, $C=-0.952$.
The estimated values for $T_1$ and $T_2$ change very little
if we choose different random realizations for the initial state
of the environment or for the model parameters  
$\Omega _{i,j}^{\alpha }$ and $\Delta _{i,j}^{\alpha }$
(data not shown) but if we change their range, $T_1$ and $T_2$ also change, 
as naively expected.

\begin{figure}[t]
\begin{center}
\includegraphics[width=15cm]{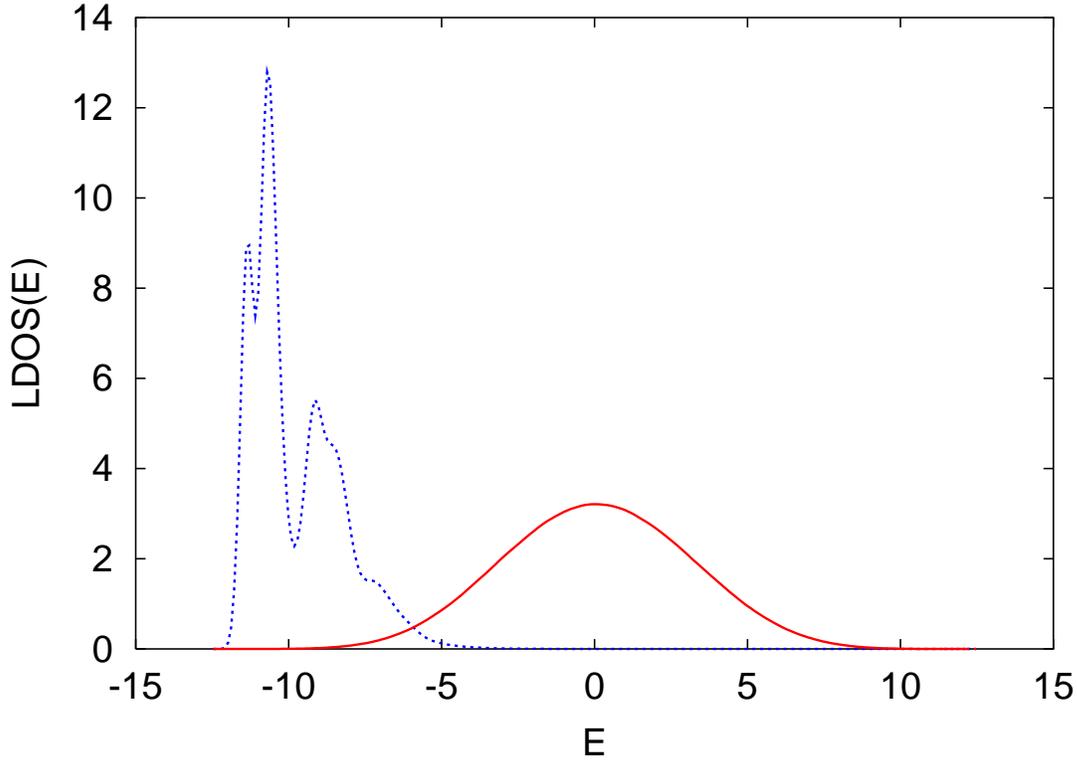}
\end{center}
\caption {(Color online)
Simulation results for the local density of states as a function
of the energy.
Solid line: Case corresponding to Fig.~\ref{fig2}\textbf{b}.
The initial state is $\vert UD\rangle _{S}\otimes \vert RANDOM\rangle _{E}$;
Dashed line: Case corresponding to Fig.~\ref{fig2}\textbf{f}.
The initial state is $\vert UD\rangle _{S}\otimes \vert GROUND\rangle _{E}$. 
}
\label{fig3}
\end{figure}

\begin{figure}[t]
\begin{center}
\includegraphics[width=15cm]{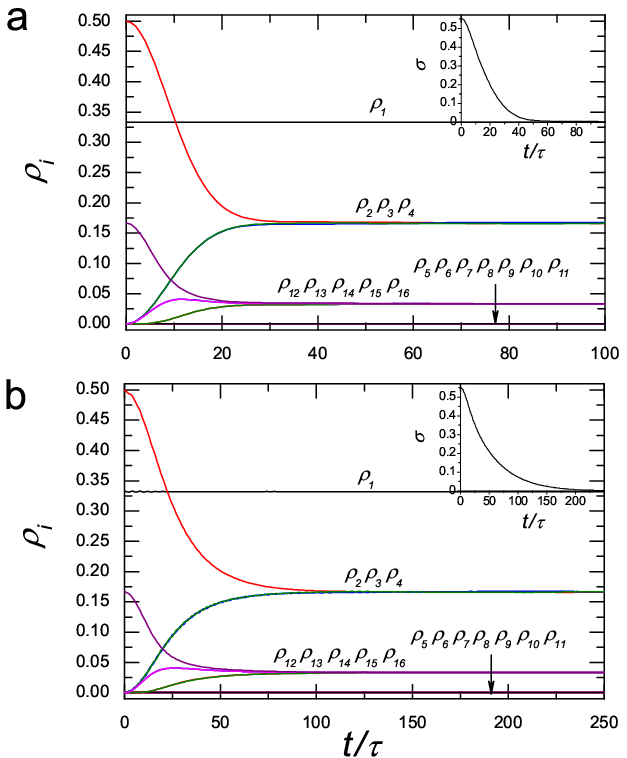}
\end{center}
\caption {(Color online)
Simulation results for a Heisenberg-ring $H_{S}$ ($J=-5$, $n_{S}=4$, initial state $\vert UD\rangle_S$)
coupled to a spin glass $H_{E}$ ($\Omega =0.15$, $n=16$,
initial state $\vert RANDOM\rangle_E $) via (\textbf{a}) Heisenberg $H_{SE}$ ($\Delta =0.075$) or
(\textbf{b}) Heisenberg-type $H_{SE}$ ($\Delta =0.15$).
Although full decoherence is observed in both cases, the
the system $S$ only relaxes to a state with equal probabilities within
each energy subspace, that is to  a ``microcanonical'' state per energy subspace.}
\label{fig4}
\end{figure}

The simulation demonstrates that the system
first looses all coherence and then, on a longer time-scale,
relaxes to its thermal equilibrium state with a finite
temperature. In terms of the theory of magnetic resonance~\cite{Abragam61},
$T_{1}$ and $T_{2}$ are the times of dissipation and dephasing, respectively. Note that
in contrast to the cases considered in the theory of nuclear magnetic
resonance, in most of our simulations, $H_{S}$, $H_{E}$ and $H_{SE}$ are comparable so
the standard perturbation derivation of $\sigma$ and $E$ does not work.
In the case of very small $H_{E}$, one should expect,
instead of an exponential decay of $\sigma$ and $E$, a Gaussian decay,
as observed in our earlier work~\cite{Yuan2006,Yuan2007,Yuan2008}.

Results for systems ($H_S$) with different symmetries and connectivities
that interaction with the same type of environments ($H_E$) via the same type
of couplings ($H_{SE}$) are shown in Fig.~\ref{fig2}.
The systems used are an XY-ring, a Heisenberg-ring,
an Ising-ring, a Heisenberg-triangular-lattice, and a spin glass.
From Fig.~\ref{fig2}, it is clear that
independent the internal symmetries and connectivity of the system 
and independent the initial state of the whole system 
(except for case \textbf{f} in which the environment is initially in its ground state),
all systems relax to a state with full decoherence.
Notice that in case \textbf{b}, $\sigma$ vanishes exponentially with time,
whereas in other cases (\textbf{a},\textbf{c},\textbf{d},\textbf{e}), $\sigma$ initially increases
and then vanishes exponentially with time, due to the 
entanglement between the system and the environment.
This observation is in concert with our earlier work~\cite{Yuan2006,Yuan2007,Yuan2008}.

Furthermore, in all cases except \textbf{f}, 
the system always relaxes to a canonical distribution 
($\delta\rightarrow 0$) as soon as it is in the state with full decoherence ($\sigma\rightarrow 0$),
indicating that the time of decoherence ($T_2$) and the thermalization is almost the same. 
In agreement with the results depicted in Fig.~\ref{fig1}, the decoherence time $T_2$
is shorter than the typical time scale $T_1$ on which the system and environment
exchange energy and the effective inverse temperature $b(t)$ reaches its stationary value.

The case \textbf{f} is easily understood in terms of the local density of states.
In Fig.~\ref{fig3} we show the LDOS for the cases \textbf{b} and \textbf{f}, the only difference
between these two cases being the initial state of the environment.
Up to a trivial normalization factor, the LDOS curve for case \textbf{b} 
is indistinghuisable from the density of states
(data not shown) calculated from the solution of the TDSE using the technique described in Ref.~\cite{HAMS00}
This suggests that if the environment starts from the random superposition of all its states,
all states of the whole system may participate in the decoherence/relaxation process.
In contrast, the LDOS curve for case \textbf{f} has a very small overlap with the 
density of states (the curve of which coincides with the solid line in Fig.~\ref{fig3}).
Therefore, starting with an environment in the ground state, 
only a relatively small number states participates in the decoherence process,
as confirmed by the results for $\sigma(t)$ shown in Fig.~{\ref{fig2}}\textbf{f}.

For completeness, we discuss a two other situations in which, for fairly obvious reasons, 
the system cannot relax to its canonical distribution.
Obviously, if the energy of the system is conserved ($[H_{S},H]=0$), the system cannot exchange energy with the environment
and we should not expect relaxation to the canonical distribution.
In this case, as shown in Fig.~\ref{fig4}, after the system $S$ 
has reached a state with full decoherence,
its density matrix does not converge to the canonical state.
Likewise, if the range of energies of the environment $E$ is
too small compared to that of the system ($\vert \Omega \vert \ll \vert J\vert$)
as in the example shown in Fig.~\ref{fig4}b,
there is no convergence to the canonical state either.
It is to be noted that in both cases, the interaction with the environment
leads to perfect decoherence ($\sigma(t)\approx0$, see insets)
such that the reduced density matrix converges to a diagonal matrix.
However, from Fig.~\ref{fig4}, it follows that $S$ relaxes to 
a kind of microcanical state in which the states in each energy subspace
have equal probability, the probabilities to end up in a subspace 
depending on the initial state.

Disregarding the three cases mentioned earlier, 
the simulation results presented in Figs.~(\ref{fig1}) and (\ref{fig2})
suggest that the state of a system generally relaxes to the canonical distribution when
the system is coupled to an environment of which the dynamics is sufficiently complex
also in the case that the interaction between system and environment cannot be regarded as a perturbation.
There are exceptions but these are easily understood: Either there are not enough states available 
for the decoherence (Fig.~{\ref{fig2}}\textbf{f}) to yield a diagonal
reduced density matrix or the energy relaxation (Fig.~{\ref{fig4}})
is not effective in letting the diagonal reduced density matrix relax to the canonical distribution.
 
Although we have only presented results for a spin glass 
environment $H_{E}$, our results (not shown) for any of the choices for
$H_{S}$ and $H_{E}$ mentioned earlier, in combination 
a Heisenberg-type $H_{SE}$ interaction between system and environment,
or for $H_{S}$ and $H_{SE}$ in combination Heisenberg-type $H_{E}$
leads to the same conclusion, namely that the state of a system relaxes the canonical distribution.

\section{Discussion}

The results presented here have been obtained from an \emph{ab initio} numerical solution of the TDSE
in the absence of, for instance, dissipative mechanisms,
and demonstrate that the existence of the canonical distribution, 
a basic postulate of statistical mechanics, is a direct consequence of quantum dynamics.

We have shown that if we have a system $S$ that interacts with an environment $E$
and the whole system $S+E$ forms a closed quantum system that evolves in time according to the TDSE,
$S$ and $E$ can exchange energy, the range of energies of 
$E$ is large compared to the range of energies of $S$,
and the interaction between $S$ and $E$ leads to full decoherence of $S$,
then the state of $S$ relaxes to the canonical distribution.
Note that only the condition of full decoherence is a nontrivial requirement.

We emphasize that our conclusion does not rely on time averaging of observables,
in concert with the fact that real measurements of 
thermodynamic properties yield instantaneous, not time-averaged, values.
Furthermore and perhaps a little counter intuitive, our results show that
relatively small environments ($\approx20$ spins) are sufficient to 
drive the system $S$ to thermal equilibrium and that there is
no need to assume that the interaction between the system and environment
is weak, as is usually done in kinetic theory.

In conclusion: The work presented here strongly suggests that the canonical ensemble, being
one of the basic postulates of statistical mechanics, 
is a natural consequence of the dynamical evolution of a quantum system.
This conclusion may be exciting but as quantum mechanics describes 
the dynamics of a system and statistical mechanics gives us 
the distribution when the system is in the equilibrium state,
these two successful theories should not be in conflict 
once the conditions for the system to relax to its thermal equilibrium 
are satisfied.

\section{Aknowledgments}
It is a pleasure to thank S. Miyashita, F. Jin, S. Zhao and K. Michielsen for many helpful discussions. 
We are grateful to S. Miyashita and M. Novotny for several suggestions to improve the manuscript.
This work was partially supported by NCF, The Netherlands.

\end{document}